\documentclass{pasj00}

\begin{document}
\SetRunningHead{Y. Takeda et al.}{Potassium Abundances in Red Giants 
of Globular Clusters}
\Received{2009/01/21}
\Accepted{2009/02/23}

\title{Potassium Abundances in Red Giants of \\
Mildly to Very Metal-Poor Globular Clusters
\thanks{Based on data collected at Subaru Telescope, which is operated
by the National Astronomical Observatory of Japan.}
\thanks{The electronic table E1 will be made available 
at the PASJ web site upon publication, while it is provisionally placed at 
$\langle$http://optik2.mtk.nao.ac.jp/\~{ }takeda/GCpotassium/$\rangle$.}
}

%

\author{
Yoichi \textsc{Takeda,}\altaffilmark{1,2}
Hiroyuki \textsc{Kaneko,}\altaffilmark{2}
Naoko \textsc{Matsumoto,}\altaffilmark{2}\\
Shoichi \textsc{Oshino,}\altaffilmark{2}
Hiroko \textsc{Ito,}\altaffilmark{2}
and
Takatoshi \textsc{Shibuya}\altaffilmark{2}
}

\altaffiltext{1}{National Astronomical Observatory, 2-21-1 Osawa, 
Mitaka, Tokyo 181-8588}
\email{takeda.yoichi@nao.ac.jp}
\altaffiltext{2}{The Graduate University for Advanced Studies, 
2-21-1 Osawa, Mitaka, Tokyo 181-8588}
\email{kaneko-h@nro.nao.ac.jp, naoko.matsumoto@nao.ac.jp, 
shoichi.oshino@nao.ac.jp,\\ 
hiroko.ito@nao.ac.jp,  takatoshi.shibuya@nao.ac.jp}
%


\maketitle

\begin{abstract}
A non-LTE analysis of K~{\sc i} resonance lines at 7664.91 and 
7698.97~$\rm\AA$ was carried out for 15 red giants belonging
to three globular clusters of different metallicity (M~4, 
M~13, and M~15) along with two reference early-K giants
($\rho$~Boo and $\alpha$~Boo), in order to check whether
the K abundances are uniform within a cluster and to investigate 
the behavior of [K/Fe] ratio at the relevant metallicity range 
of $-2.5 \ltsim$~[Fe/H]~$\ltsim -1$.
We confirmed that [K/H] (as well as [Fe/H]) is almost homogeneous 
within each cluster to a precision of $\ltsim 0.1$~dex, though 
dubiously large deviations are exceptionally seen for two 
peculiar stars showing signs of considerably increased turbulence 
in the upper atmosphere.
The resulting [K/Fe] ratios are mildly supersolar by a few tenths
of dex for three clusters, tending to gradually increase from 
$\sim$~+0.1--0.2 at [Fe/H]~$\sim -1$ to $\sim +0.3$ at 
[Fe/H]~$\sim -2.5$. This result connects reasonably well with the 
[K/Fe] trend of disk stars ($-1 \ltsim$ [Fe/H]) and that of 
extremely metal-poor stars ($-4 \ltsim$~[Fe/H]~$\ltsim -2.5$).
That is, [K/Fe] appears to continue a gradual increase from 
[Fe/H] $\sim 0$ toward a lower metallicity regime down to 
[Fe/H]~$\sim -3$, where a broad maximum of [K/Fe]~$\sim$ +0.3--0.4 
is attained, possibly followed by a slight downturn at 
[Fe/H]~$\ltsim -3$.
\end{abstract}

%


\section{Introduction}

The chemical evolution of potassium (K) in the Galaxy,
which can be investigated by examining its photospheric abundances 
of old metal-poor stars, is still only insufficiently explored and 
not yet well understood, especially in the metal-poor regime
of halo stars. The main reason for this unsatisfactory situation
is presumably the difficulty involved with determinations of
stellar potassium abundances, for which only two strong K~{\sc i} 
lines at 7664.91 and 7698.97~$\rm\AA$ (hereinafter referred 
to as 7665 and 7699 lines for brevity) are practically usable.
That is, although these resonance lines are so strong as to be
measurable even for stars of extremely low metallicity,
the following features make the problem comparatively 
harder to deal with:\\
--- Above all things, they generally suffer an appreciably large non-LTE 
effect,\footnote{While non-LTE line formation of K~{\sc i} 7665/7699 
resonance lines specific to the solar atmosphere had begun already in 1980s 
(see Bruls et al. 1992 and the references therein), non-LTE studies
of these K~{\sc i} doublet lines directed to stars other than the Sun 
gradually appeared in the last decade: Takeda et al. (1996)
carried out a detailed study on the formation of this line in Procyon 
(along with the Sun). Ivanova and Shimanski\u{\i} (2000) calculated
non-LTE abundance corrections applicable to A--K stars of wide 
parameter ranges. Similarly, Takeda et al. (2002a) carried out extensive 
non-LTE calculations on a grid of models and published tables of 
non-LTE corrections applicable to F--G--K dwarfs through supergiants.
Recently, Zhang et al. (2006a) performed a careful non-LTE investigation 
of solar potassium lines with a special attention to clarifying the 
important atomic parameter of neutral hydrogen collision cross section, 
which they further applied to determinations of K abundances for 
metal-poor stars (Zhang \& Zhao 2005, Zhang et al. 2006b).}
which depends on the stellar atmospheric parameters as well
as the line strength. Actually, since the extent of the (negative) 
non-LTE correction amounts from a few tenths dex even up to 
$\sim 1$~dex, it is requisite to take account of the non-LTE effect
in deriving stellar K abundances. \\
--- Besides, they occasionally suffer serious blending with 
strong absorption lines originated from earth's atmosphere (especially 
for the 7665 line) because of being located in an unfavorable 
wavelength region crowded with such telluric lines. Hence, when 
encountered with such unfortunate cases, one has to recover the pure
stellar spectrum by appropriately eliminating the blended telluric
spectrum.

The first non-LTE study of K abundances in metal-poor stars was 
carried out by Takeda et al. (2002a, hereinafter referred to as
Paper I). They showed that [K/Fe] values for the main-sequence stars 
in the galactic disk ($-1 \ltsim$~[Fe/H]~$\ltsim 0$) tend to show 
a rather tight relation of steadily increasing with a decrease of 
metallicity (such as like the trend of $\alpha$-group elements).
This was actually a reconfirmation of the results of Chen et al.'s 
(2000) LTE analysis, despite the significant non-LTE corrections of 
$\sim -0.5$~dex, because the corrections turned out to act similarly
on the Sun and these disk dwarfs (i.e., [K/H], the differential 
stellar potassium abundance relative to the Sun, was eventually 
not much affected). Such an $\alpha$-like trend of [K/Fe] for disk 
stars was also confirmed by Zhang et al. (2006b).

Unfortunately, the situation becomes quite uncertain when we go 
into the much lower metallicity regime ([Fe/H]~$\ltsim -1$).
It was already noticed in Paper I from the non-LTE reanalysis of 
the published data for several halo stars of [Fe/H] $\sim -2$
(Gratton \& Sneden 1987b) that their [K/Fe] values widely spread 
from $\sim$ +0.1 to $\sim$ +0.7 (cf. figure 4a of Paper I; see also 
figure 8 of Gratton \& Sneden 1987a). Zhang and Zhao (2005) also 
suggested a considerably large diversity of [K/Fe] (from $\sim 0$ 
to $\sim 1$) for the halo stars of $-2 \ltsim$~[Fe/H]~$\ltsim -1$ 
they studied (cf. figure 12 therein). On the other hand, however, 
five halo stars included in Zhang et el.'s (2006) study have 
nearly the same [K/Fe] ratios around $\sim +0.2$ over the range of 
$-2 \ltsim$~[Fe/H]~$\ltsim -1$ (cf. figure 5 therein). 
Besides, Cayrel et al. (2004) 
obtained in their extensive study of extremely metal-poor stars 
($-4 \ltsim$~[Fe/H]~$\ltsim -2.5$) that [K/Fe] shows a slightly 
supersolar trend ($\sim$ +0.1--0.2) decreasing toward a very low 
metallicity with a fairly small scatter (cf. figure 9 therein).
Thus, at present, we know little about how the [K/Fe] ratio
actually behaves itself at [Fe/H] $\ltsim -1$: Tight trend? 
Large diversity? Does it smoothly connect to the $\alpha$-like 
clear tendency seen in disk stars at [Fe/H] $\gtsim -1$?

For the purpose of checking the possibility of K abundance
diversity in such metal-deficient halo stars, it would be 
interesting to study stars in globular clusters, for which
any trial of potassium abundance determination has never 
been reported so far to our knowledge.
Admittedly, the elemental abundances of old globular cluster 
stars may not necessarily be the same as those of field halo stars, 
given that specific abundance peculiarities are known for several
elements (e.g., O, Na, Mg, Al) which are presumably due to dredge-up 
of nuclear-processed product caused by evolution-induced deep mixing.
However, since K is considered to be synthesized mainly via the 
oxygen burning in high-mass stars, whichever the process is
explosive or hydrostatic (see, e.g., table 19 in Woosley \& Weaver 1995), 
it is unlikely that the surface K abundance 
undergoes any appreciable changes during the course of stellar 
evolution in low-mass stars of globular clusters.
Besides, the fact that no star-to-star variation in globular clusters 
is observed in the abundances of Ca ($Z$~=~20, near to $Z$~=~19 of K) may 
also suggest the inertness of K to nuclear processes in the stellar 
interior, presumably because of the higher Coulomb barrier compared 
to lighter elements.
Therefore, as a reasonable working hypothesis, we may postulate
(such as the case for Fe in most clusters) that (i) the abundance 
of K was the same in any stars of a given globular cluster when they 
were born, that (ii) this uniformity has been retained up to now, 
and that (iii) the surface K abundances of cluster stars may be 
regarded to be equivalent to field halo stars of similar metallicity.
Accordingly, if we could observationally confirm postulation (ii),
this would assure the practical validity of (iii), which 
means that we may directly compare the results of cluster stars
with those of other halo stars in general. Or, alternatively,
if we found a markedly large diversity of K abundances within 
a cluster contrary to postulation (ii), we would have to 
consider a possibility of real star-to-star variation (or cast 
doubt on the validity of our abundance determination method).

Motivated by this consideration, we decided to conduct
a spectroscopic study on the intrinsically bright red-giant stars 
of three globular clusters (M~4, M~13, and M~15) covering a wide
metallicity span (from [Fe/H] $\sim -2.5$ to $\sim -1$)
based on the high-dispersion spectra obtained with Subaru/HDS,
in order to determine the abundance of K for each star
while taking into account the non-LTE effect, after having
established the atmospheric parameters including [Fe/H].
What we intend to clarify is, as described above, to check the 
homogeneity of [K/H] within a cluster [postulation (ii)], and to study 
the behavior of [K/Fe] in halo stars based on the results of these 
globular clusters [postulation (iii)] while comparing with those 
of other metallicity regime. This is the purpose of this study.

Besides, in connection with the main subject, we carried out a 
reanalysis of the equivalent-width data of K~{\sc i} 7665 and 
7669 lines for $\sim 30$ extremely metal-poor stars published by 
Cayrel et al. (2004), in order to establish the [K/Fe] vs. [Fe/H] 
relation at the metallicity range of $-4 \ltsim$~[Fe/H]~$\ltsim -2.5$.
The main motivation is to properly take into account the non-LTE 
effect, since their treatment in this respect does not appear to 
be sufficiently valid (i.e., they applied a tentatively assumed 
correction uniformly to all stars). This reanalysis is separately 
described in the Appendix.

\section{Observational Data}

Three globular clusters (M~4, M~13, and M~15) were chosen for this 
study, because (1) they have different metallicities from each other 
([Fe/H]$_{\rm M15} < $~[Fe/H]~$_{\rm M13} < $~[Fe/H]$_{\rm M4}$),
(2) they are located in near or reasonable distance to us 
(so that red giants of $V \ltsim 12$~mag exist), and (3) they 
are comparatively well studied and a number of literature data 
may be found. Practically, we selected 5 red-giant stars for 
each cluster satisfying the criterion of 
4100~K~$\ltsim T_{\rm eff}^{\rm pho} \ltsim 4300$~K
($T_{\rm eff}^{\rm pho}$ is the effective temperature photometrically
evaluated from $(B-V)_{0}$ colors; cf. subsection 5.1):
L~2406, L~2617, L~3209, L~3624, and L~4511 for M~4; 
I-13, II-76, III-52, III-59, and III-73 for M~13;
K~144, K~341, K~431, K~634, and K~825 for M~15.
These targets belong to the brightest-class group in each cluster 
($V \sim$~11--12~mag for M~4, $V \sim$~12--13~mag for M~13 and M~15).
In addition, two early-K giant stars of near-solar or subsolar 
metallicity ($\rho$~Boo and $\alpha$~Boo) were also included as the 
standard stars, though they are normal K giants of luminosity class III 
with $\log(L/L_{\odot})\sim 2$ while all 15 cluster stars are 
``tip giants'' locating on the tip of asymptotic giant branch at 
$\log(L/L_{\odot})\sim 3$.

The observations of these 17 stars (along with a rapid rotator Altair 
as a reference of telluric lines) were carried out on the night of 
2008 August 20 (Hawaii Standard Time) by using the High Dispersion 
Spectrograph (HDS; Noguchi et al. 2002) placed at the Nasmyth 
platform of the 8.2-m Subaru Telescope, which can record 
high-dispersion spectra covering a wavelength portion of 
$\sim 2600 \rm\AA$ (in the red cross disperser mode) with 
two CCDs of 2K$\times$4K pixels at a time. 
With the slit width set at $0.''6$ (300 $\mu$m) and a binning of 
2$\times$2 pixels, the resolving power of the obtained spectra is 
$R \sim 60000$. In the standard ``Ra'' setting, our spectra cover 
the wavelength region of 5100--6400 $\rm\AA$ (blue CCD) and 
6500--7800 $\rm\AA$ (red CCD), which was so chosen as to make 
use of the yellow--red region of $\lambda \gtsim 5000 \rm\AA$ 
(where the sensitivity of CCD is large and many Fe lines usable 
for parameter determinations exist) while including the targeted
K~{\sc i} 7665/7699 lines amply.  
The seeing size was $0.''5$--$0.''6$, and all of the targets
could be successfully observed without any significant influence
of neighboring stars.
The integrated exposure time for each cluster star was from 10~min 
(= 5~min $\times 2$; for M~4 stars) to 15--20~min 
(= 7.5--10~min $\times 2$; for M~13 or M~15 stars).

The reduction of the spectra (bias subtraction, flat-fielding, 
scattered-light subtraction, spectrum extraction, wavelength calibration,
continuum normalization) was performed by using the {\tt echelle} 
package of the software IRAF\footnote{IRAF is distributed
    by the National Optical Astronomy Observatories,
    which is operated by the Association of Universities for Research
    in Astronomy, Inc. under cooperative agreement with
    the National Science Foundation.} 
in a standard manner. The resulting average S/N ratios were around 
$\sim$~100--150 for each of the 15 cluster stars, while much 
higher values were attained for the bright reference stars;
$\sim$~300--400 ($\rho$~Boo), $\sim$~200--300 ($\alpha$~Boo),
and $\sim$~600--700 (Altair).

\section{Atmospheric Parameters}

The determination of the atmospheric parameters [$T_{\rm eff}$ 
(effective temperature), $\log g$ (surface gravity), $v_{\rm t}$ 
(microturbulence), and [Fe/H] ($\equiv A_{\rm Fe}^{\rm star} - 
A_{\rm Fe}^{\odot}$; differential Fe abundance relative to 
the Sun, where $A_{\rm Fe}^{\odot}$ is 7.50 in the usual 
normalization of $A_{\rm H} =12.00$)]  
necessary for constructing model atmospheres was implemented  
by way of the spectroscopic approach using the equivalent widths
($EW$s) of Fe~{\sc i} and Fe~{\sc ii} lines, which has a merit of 
establishing these four parameters based only on the same 
spectrum to be further used for abundance determinations. 

Practically, we used the computer program (named TGVIT) developed 
for this purpose (Takeda et al. 2005; cf. section 2 therein),
which is based on the principle of searching for the most 
optimum solution in the 3-dimensional ($T_{\rm eff}$, $\log g$, 
$v_{\rm t}$) space such that simultaneously satisfying the three 
requirements: (1) $\chi_{\rm low}$-independence of Fe~{\sc i} 
abundances (excitation equilibrium, where $\chi_{\rm low}$ is the
lower excitation potential), (2) equality of the mean abundance
derived from Fe~{\sc i} lines and that from Fe~{\sc ii} lines 
(ionization equilibrium), and (3) $EW$-independence of the 
abundances (curve-of-growth matching), as described in Takeda, Ohkubo,
and Sadakane (2002b). Since low-gravity K-type giants are 
specifically involved in this study, we newly computed 
a grid of data files [covering the parameter ranges of 3750--5000~K 
in $T_{\rm eff}$ (K), 0.0--2.5 in $\log g$ (cm~s$^{-2}$), 0.0--4.0 
in $v_{\rm t}$ (km~s$^{-1}$), and $-2.5$ to +0.1 in [Fe/H] (dex)] 
to be used in application of TGVIT by interpolation (or extrapolation). 

First, we measured the $EW$s of available lines by consulting the 
list of 330 Fe lines (cf. electronic table E1 in Takeda et al. 2005)
by using the Gaussian fitting method, where only lines weaker than 
100~m$\rm\AA$ were used in order to make sure that errors caused 
by damping wings/parameters are suppressed to a negligible level. 
Then, given these $EW$ data as inputs, the converged solutions of 
$T_{\rm eff}$, $\log g$, and $v_{\rm t}$  (along with $A_{\rm Fe}$ 
as a product) were established by iteratively running TGVIT
(cf. subsection 3.2 in Takeda et al. 2005).
The resulting parameter solutions are summarized in table 1,
while the detailed $EW$ data and the Fe abundances corresponding to 
the final parameters for each star are presented in electronic table E1.
The trend of the Fe abundances corresponding to the final solutions 
is plotted against $EW$ as well as $\chi_{\rm low}$ in figure 1, 
where we can see that the required conditions are reasonably 
accomplished.

\begin{figure}
  \begin{center}
    \FigureFile(75mm,80mm){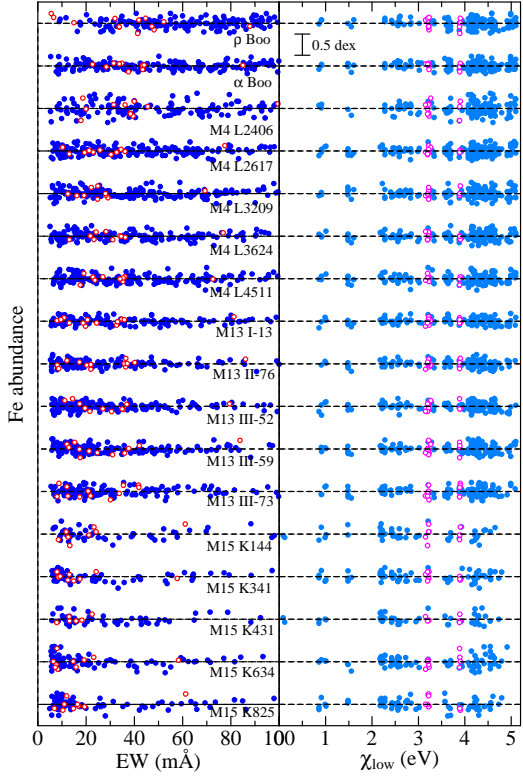}
  \end{center}
\caption{
Fe abundance vs. equivalent width relation (left panel) 
and Fe abundance vs. lower excitation potential relation
(right panel) corresponding to the finally established
atmospheric parameters of $T_{\rm eff}$, $\log g$, and $v_{\rm t}$
for each of the 17 stars. The filled (blue) and open (red) circles 
correspond to Fe~{\sc i} and Fe~{\sc ii} lines, respectively. 
The results for each star are shown relative to the mean abundance 
indicated by the horizontal dashed line, and vertically offset by 1.0 
relative to the adjacent ones. 
}
\end{figure}

The internal statistical errors involved with these solutions of 
$T_{\rm eff}$, $\log g$, $v_{\rm t}$, and [Fe/H], which were 
derived by the procedure described in subsection 5.2 of 
Takeda, Ohkubo, and Sadakane (2002b), turned out to be 
$\sim$~10--30~K, $\sim$~0.05--0.1~dex,
$\sim$~0.1--0.4~km~s$^{-1}$, and $\sim$ 0.03--0.1 dex, respectively. 
We will discuss in subsection 4.1 these spectroscopic parameters 
in comparison with those derived from the conventional method 
(i.e., use of photometric colors or evolutionary tracks). 

\section{Analysis of Potassium Lines}

\subsection{EW Measurement}

Among our 17 target stars, 5 stars of M~13 were found to be 
associated with the most unfortunate case. The K~{\sc i} 7699 
line could not be measured at all because it happened to 
fall just on the narrow inter-order gap of HDS (at 
7687--7694~$\rm\AA$) due to M~13's large (negative) 
radial velocity of $\sim -220$ to $-230$~km~s$^{-1}$, 
while the K~{\sc i} 7665 line was seriously blended with the 
strong telluric O$_{2}$ line at 7659.3~$\rm\AA$. Such a 
contamination in the K~{\sc i} 7665 region is also seen for 
the case of two standard stars, $\alpha$~Boo and (especially) 
$\rho$~Boo, which turned out to be influenced by the telluric 
O$_{2}$ doublet at $\sim$~7664/7665~$\rm\AA$.
We, therefore, tried to eliminate the effect of telluric lines 
for these 7 stars by dividing the raw spectra by the spectrum
of Altair (rapid rotator) by using the IRAF task {\tt telluric}. 
The resulting as well as the original spectra are shown
in figure 2, where we can see that this elimination surely 
worked (though not perfectly). Figure 3 displays the spectra 
used for measurements of $EW_{7665}$ and $EW_{7699}$, which were
measured by the Gaussian fitting method as for the case of Fe lines.
The finally adopted values of $EW_{7665}$\footnote{Actually, 
we had to apply further corrections to $EW_{7665}$
for five M~13 stars (I-13, II-76, III-52, III-59, and III-73)
and $\rho$~Boo, for which we tried elimination of telluric lines 
($EW_{7665}$ for $\alpha$~Boo does not need to be corrected, 
since the important line core was free from contamination). 
Namely, given that this removal could not be completely done, 
as can be recognized from the appreciable residual at the position 
of the redward component of the relevant O$_{2}$ doublet 
(at 7660.4~$\rm\AA$ for M~13 stars and at 7665.9~$\rm\AA$ for 
$\rho$~Boo; see the depression marked with ``x'' in figure 2), 
we evaluated its equivalent width ($\Delta EW_{\rm resid}$)
and subtracted from the directly measured $EW$ as 
$EW_{7665}^{\rm adopt} = EW_{7665}^{\rm raw} - \Delta EW_{\rm resid}$
for these six stars, assuming that the residuals of both telluric 
components are almost the same. The applied $\Delta EW_{\rm resid}$ 
corrections are given in the caption of table 1.
} and $EW_{7699}$ are summarized in table 1.

\begin{figure}
  \begin{center}
    \FigureFile(75mm,80mm){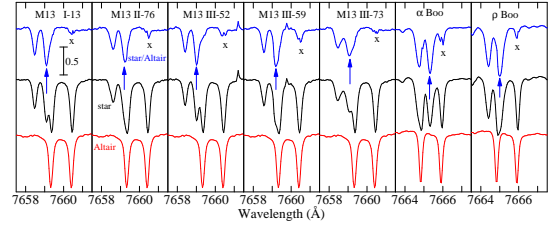}
  \end{center}
\caption{
Removal of telluric O$_{2}$ lines in the K~{\sc i} 7665 line 
region for the seven stars significantly suffering this effect. 
Dividing the raw stellar spectrum (middle, black) by the spectrum 
of a rapid rotator Altair (bottom, red) results in the final 
spectrum (upper, blue) where the effect of the blending component 
is reasonably (even if not completely) eliminated. 
The residual caused by the imperfect elimination for the redward 
component of the telluric doublet lines is marked by a cross (x), 
the strength of which is usable for the correction of 
$EW$(K~{\sc i} 7665) similarly affected by the incomplete 
removal of the blueward compenent.
The position of the K~{\sc i} 7665 line is indicated by an arrow. 
Each spectrum (normalized with respect to the continuum)
is vertically offset by 1.0 relative to the adjacent one.
No Doppler correction has been applied to the wavelength scale.
}
\end{figure}

\begin{figure}
  \begin{center}
    \FigureFile(75mm,80mm){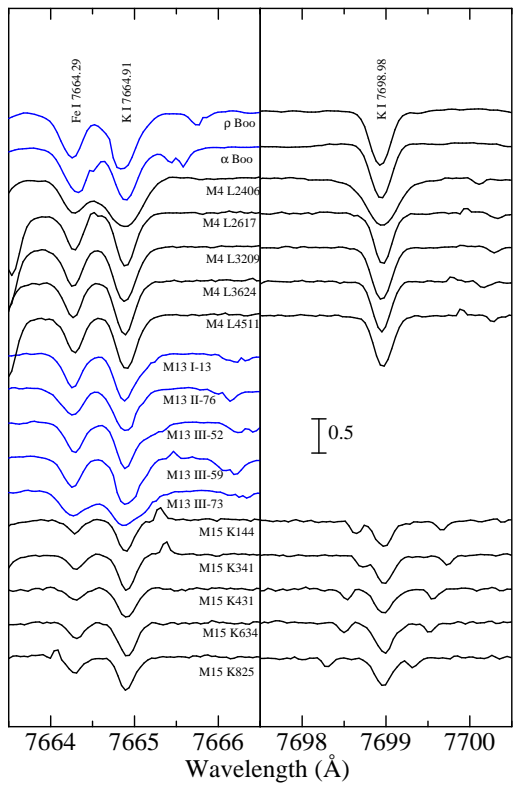}
  \end{center}
\caption{
Spectra of 17 stars in the neighborhood of the K~{\sc i} 7665 
line (left panel) and the K~{\sc i} 7699 line (right panel).
The seven spectra for the two standard stars and five
M~13 stars in the left panel (drawn in blue lines)
are the ones corrected for the telluric lines (cf. figure 2).
Each spectrum (normalized with respect to the continuum)
is vertically offset by 0.5 relative to the adjacent one.
An appropriate Doppler correction has been applied to
each spectrum so that the positions of stellar spectral lines
are located at the corresponding laboratory wavelengths.
}
\end{figure}

\subsection{Abundance Determination}

The determination of potassium abundances from the measured
$EW_{7665}$ and $EW_{7699}$ by taking into account the non-LTE
effect was carried out in almost the same manner as in Paper~I. 
As the basic grid of model atmospheres, we used Kurucz's (1993) 
ATLAS9 models (corresponding to $v_{\rm t}$ = 2~km~s$^{-1}$),
which were three-dimensionally interpolated (or extrapolated in
special cases of negative $\log g$) with respect to 
$T_{\rm eff}$, $\log g$, and [Fe/H] to generate the atmospheric
model for each star.

The non-LTE statistical-equilibrium calculations were implemented 
for a grid of 90 models resulting from combinations of three 
$T_{\rm eff}$ values (4000, 4250, 4500~K), five $\log g$ 
values (0.0, 0.5, 1.0, 1.5, 2.0), and six [Fe/H] values 
($-2.5$, $-2.0$, $-1.5$, $-1.0$, $-0.5$, 0.0), so that
we can obtain the depth-dependent non-LTE departure coefficients
for any star by interpolating (or extrapolating) this grid.
See Takeda et al. (1996) for the computational details.
We suppressed the effect of neutral-hydrogen collisions
to a negligible level by multiplying the classical value
based on Drawin's cross section (cf. Steenbock \& Holweger 1984) 
by a factor of 10$^{-4}$ according to the conclusion of that paper.

The derivation of the K abundance from a given $EW$ was done 
as in Paper I with Kurucz's (1993) WIDTH9 program, which had been 
considerably modified in various respects (especially to include 
the effect of departure from LTE). According to Takeda et al. (1996), 
the van der Waals damping parameter ($C_{6}$) was increased by 
applying a correction of $\Delta\log C_{6} = +1.00$ to the 
conventional Uns\"{o}ld's (1955) formula value (corresponding to 
using $\Gamma_{6} = 2.5 \Gamma_{6}^{\rm Unsold}$). We adopted the 
$\log gf$ values of +0.13 (7665 line) and $-0.17$ (7699 line), 
and the radiation damping constant of 
$\Gamma_{\rm rad} = 0.38\times 10^{8}$~s$^{-1}$,
which were taken from Kurucz and Bell's (1995) compilation.

Actually, we prepared two sets of non-LTE departure coefficients 
(grid of 90 models) corresponding to two different choices of input 
K abundances ([K/Fe] = 0.0 and [K/Fe] = +0.5), considering that the 
resulting [K/Fe] is likely to be encompassed by these two values. 
And two kinds of non-LTE abundances ($A_{\rm K}^{0.0}$ and $A_{\rm K}^{0.5}$) 
were obtained for a given $EW$ for each of the two sets. Then, these
$A_{\rm K}^{0.0}$ and $A_{\rm K}^{0.5}$ were interpolated (or 
extrapolated) so that the final non-LTE solution ($A_{\rm K}$) and 
the used departure coefficient becomes consistent with each other 
(cf. subsection 4.2 in Takeda \& Takada-Hidai 1994). In addition, 
we also derived the LTE abundance $A_{\rm K}^{\rm LTE}$, from which
the non-LTE correction was derived as $\Delta^{\rm NLTE} \equiv
A_{\rm K} - A_{\rm K}^{\rm LTE}$. Finally, the differential K 
abundance relative to the Sun was computed as 
[K/H] $\equiv A_{\rm K} -5.12$, where Anders and Grevesse's (1989) 
$A_{\rm K,\odot}$ value of 5.12 was adopted as the solar potassium 
abundance (which is also consistent with the result of Takeda et al. 1996).
The resulting values of [K/H] and $\Delta^{\rm NLTE}$ for each line
along with the [K/Fe] ratio ($\equiv$ [K/H] $-$ [Fe/H]; [K/H] is 
the average of two lines) are summarized in table 1.
Regarding the typical errors in the K abundances caused by 
uncertainties in the atmospheric parameters or the damping constant,
table 1 in Paper I (especially for the cases of metal-poor K giants
such as HD~122563, HD~165195, and HD~221170) may be informative.

\section{Discussion}

\subsection{Verifying Model Atmosphere Parameters}

It may be worth comparing our spectroscopically determined 
atmospheric parameters with those by another method occasionally 
used for globular cluster stars; i.e., the photometric 
$T_{\rm eff}^{\rm pho}$ and evolutionary $\log g^{\rm evo}$, where
$T_{\rm eff}$ is derived from colors and $g$ is evaluated from $L$ 
(the stellar luminosity estimated from the apparent magnitude and 
the distance), $T_{\rm eff}$, and $M$ (the stellar mass often 
assumed to be 0.8~$M_{\odot}$ for globular cluster giants) by 
the relation of $g/g_{\odot} = (M/M_{\odot}) (L/L_{\odot})^{-1} 
(T_{\rm eff}/T_{\rm eff, \odot})^{4}$.

The basic data of $V$ magnitude and $B-V$ color were taken from 
Ivans et al. (1999; M 4), Smith and Briley (2006; M 13), 
and Sneden et al. (1997; M 15); the color excess $E_{B-V}$ and 
the distance $d$ (assumed to be the same for all targets in a cluser) 
were taken from the on-line database\footnote
{$\langle$http://dipastro.pd.astro.it/globulars/$\rangle$.}
elaborated by the Padova group. For $\rho$~Boo and $\alpha$~Boo,
$E_{B-V}$ was assumed to be zero, and the other data were taken 
from the SIMBAD database.
We first derived $T_{\rm eff}$ from the dereddened $(B-V)_{0}$
color by using Alonso, Arribas, and Mart\'{\i}nez-Roger's (1999) equation (4) in their
table 2, where $0.0$ ($\rho$~Boo), $-0.6$ ($\alpha$~Boo), 
$-1.2$ (M~4 stars), $-1.7$ (M~13 stars), and $-2.5$ (M~15 stars)
were assumed for [Fe/H]. Then, since the stellar luminosity $L$ can 
be obtained from the extinction-corrected $V_{0}$ magnitude 
($V - 3.1 E_{B-V}$), the distance $d$, and the bolometric correction
evaluated by Alonso, Arribas, and Mart\'{\i}nez-Roger's (1999) equations (17) and (18),
we can derive $\log g^{\rm evo}$ from the above-mentioned relation
by assuming $M = 0.8 M_{\odot}$. The resulting $T_{\rm eff}^{\rm pho}$ 
and $\log g^{\rm evo}$ (along with the corresponding [Fe/H] values 
when these parameters were used; cf. table 3) are summarized
in table 2, where the published values taken from 
various literature (since 1980's) are also presented for comparison.
Besides, table 3 gives the differences of these parameters 
relative to the adopted spectroscopic ones, and the resulting
variations in the abundances of Fe as well as K caused by
these parameter changes.

We can see the following tendency from an inspection of table 3:\\
--- In the mildly metal-poor (M~4) or near-solar metallicity 
($\alpha$~Boo, $\rho$~Boo) region, we can not see any remarkably
systematic difference in $\delta T_{\rm eff} (\equiv 
T_{\rm eff}^{\rm pho} - T_{\rm eff}^{\rm spe})$ or 
in $\delta \log g (\equiv \log g^{\rm evo} - \log g^{\rm spe})$; 
i.e., $\delta T_{\rm eff} \sim \pm$~100~K and 
$\delta \log g \sim \pm$~0.2--0.3~dex.\\
--- However, in the very metal-poor regime of [Fe/H]~$\ltsim -1.5$
(M~13 and M~15), the inequality trend of $\delta T_{\rm eff} > 0$ 
as well as $\delta \log g > 0$ manifestly appears, especially for
the most metal-poor case of M~15 where 
$T_{\rm eff}^{\rm spe}$/$\log g^{\rm spe}$ is appreciably lower than
$T_{\rm eff}^{\rm pho}$/$\log g^{\rm evo}$ by 
$\sim$~100--200~K/$\sim$~1~dex (even $\log g^{\rm spe} <0$ cases
are seen, which is hardly possible for hydrostatic stellar 
atmospheres).

Given the existence of such an appreciable discrepancy, we tend 
to wonder whether our use of spectroscopic parameters is really 
justified. Which sets ($T_{\rm eff}^{\rm spe}$/$\log g^{\rm spe}$ 
vs. $T_{\rm eff}^{\rm pho}$/$\log g^{\rm evo}$) should be 
preferably used for abundance determinations?  
Before going into this, we point out that that our targets of
intrinsically bright red giants are on or near to the AGB tip
where stars (in their late stage of evolution) are likely 
to have extended envelopes and show significant mass loss 
as well as time variability.
Actually, as demonstrated in figure 4, not a few stars show 
prominent emissions and blue-shifted cores in H$\alpha$ 
(especially, emissions are seen in all five M~15 stars), 
indicating the existence of such active phenomena.

\begin{figure}
  \begin{center}
    \FigureFile(75mm,80mm){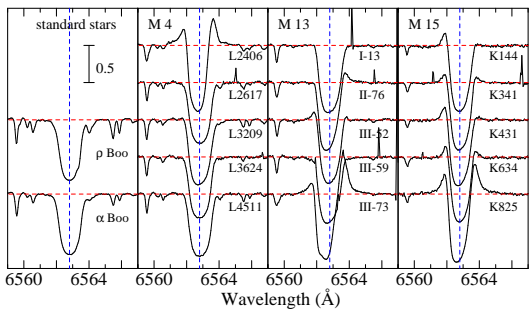}
  \end{center}
\caption{Spectra of the 17 program stars in the 
6559--6567~$\rm\AA$ region comprising H$\alpha$.  
Each spectrum is normalized with respect to the continuum
(horizontally-drawn red dashed line) and is vertically shifted 
relative to the adjacent one by an appropriate offset (1.0 or 0.5).
The wavelength scale is adjusted to the same system as in 
figure 3 where the stellar metallic lines locate
at the laboratory wavelengths. The position of the H$\alpha$ line 
center (6562.8~$\rm\AA$) in this reference frame is indicated 
by the vertically-drawn blue dashed line.
The redward emission for III-73 (M~13) is so strong 
(peaked at $\sim 1.6$) that it is crossing the profile of 
III-59 just above.
}
\end{figure}

In such cases, we would state that the spectroscopic method 
is more advantageous:\\
(1) If stars are unstable and variable in time, parameters 
based on observational data (colors, magnitudes, etc.) 
simply taken from catalogues are not reliable any more. 
Meanwhile, spectroscopic parameters are established from 
the spectrum itself, from which abundances are derived.\\
(2) We must recall that most stellar model atmospheres widely
used are based on the assumption of hydrostatic equilibrium 
and the plane-parallel approximation. In case where these 
assumptions failed, it would not be sensible any more
to assign parameters derived from the true fundamental 
stellar quantities (e.g., $\log g^{\rm evo}$ derived from 
realistic $L$, $M$, ...). On the other hand, it is still 
possible to apply conventional model atmospheres, if we could 
choose their parameters so carefully as to reproduce
the physical condition of the real atmosphere 
(while regarding that $T_{\rm eff}$ and $\log g$ are not so much 
real physical quantities as rather adjustable parameters).
Our $T_{\rm eff}^{\rm spe}$ and $\log g^{\rm spe}$ should thus 
be interpreted in this sense.

Consequently, according to our opinion, the appreciably low
$T_{\rm eff}^{\rm spe}$ and low $\log g^{\rm spe}$ in M~13
and M~15 are nothing but a manifestation of the existence of 
extended cooler region of lower density affecting the formation of 
spectral lines. As far as we invoke classical plane-parallel 
model atmospheres, such low parameter values must have been 
assigned to reproduce the relevant atmospheric condition.
Since this kind of discordance is pronounced in M~13 
([Fe/H] $\sim -1.7$) and especially in M~15 ([Fe/H] $\sim -2.5$) 
while not in M~4 ([Fe/H] $\sim -1.1$), this effect is considered
to be metallicity-dependent, which may presumably be related to 
the observational fact pointed out by Meszaros, Dupree, and Szentgyorgyi (2008)
that the outflow velocities of red giants are higher in 
metal-poor clusters (M~15) than in metal-rich clusters (M~4). 
To sum up, we consider that the use of $T_{\rm eff}^{\rm spe}$ and 
low $\log g^{\rm spe}$ is surely reasonable and preferable to
other choices.\footnote{As an alternative interpretation, 
it may be possible to consider that the large discrepancy 
in $\log g$ (systematically low $\log g^{\rm spe}$) 
seen very metal-poor red giants is due to the non-LTE overionization
effect (Fe~{\sc i} lines are weakened compared to the case of LTE, 
while Fe~{\sc ii} lines are practically unaffected), which would naturally
yield an underestimation of $\log g^{\rm spe}$ if LTE is assumed.
We can not exclude this possibility, as Ruland et al. (1980) once 
reported a sign of non-LTE overionization in low-excitation
Fe~{\sc i} lines (while high-excitation lines are comparatively 
inert) in their analysis of the early-K giant Pollux 
($T_{\rm eff} \sim 4800$~K, $\log g \sim 2.2$, and the 
near-solar metallicity). 
On the other hand, theoretical investigations on the non-LTE effect 
in the formation of Fe lines in red giants are difficult, because 
the results are sensitively influenced by uncertainties in 
computational details (e.g., the treatment of UV photoionizing
radiation field, collisional cross section with neutral 
hydrogen atoms; see, e.g., Steenbock 1985 or Takeda 1991). 
As far as the present case is concerned, however, we consider
that the possibility of an appreciable non-LTE overionization 
in Fe~{\sc i} such as claimed by Ruland et al. (1980) is rather 
unlikely. That is, since the characteristic feature they found
was ``low-excitation Fe~{\sc i} lines tend to yield lower LTE 
abundances than high-excitation Fe~{\sc i} lines'', the LTE 
excitation temperature would appear higher than usual, by which 
an {\it overestimation} of $T_{\rm eff}^{\rm spe}$ should eventually 
result. This is, however, just the opposite to what we found 
in M~13 or M~15 stars.}

\subsection{Homogeneity of Fe and K Abundances}

The results of [Fe/H], [K/H]$_{7665}$, and [K/H]$_{7699}$ 
(cf. table 1) for each star are graphically depicted in figure 5,
from which we can read the following characteristics:\\
--- The [Fe/H] values are considered to be nearly uniform
in all three clusters (M~4, M~13, and M~15) to within a precision of 
$\sim$~0.05--0.1~dex. Actually, the mean $\langle$[Fe/H]$\rangle$
(and the standard deviation $\pm \sigma$) averaged over 5 stars are 
$-1.14 (\pm 0.04)$, $-1.65 (\pm 0.09)$, and $-2.48 (\pm 0.07)$
for M~4, M~13, and M~15, respectively.\footnote{For reference, 
a simple averaging of the literature data given in table 2 
(excluding our results) gives the mean cluster metallicties 
of $\sim -1.2$ (M~4), $\sim -1.5$ (M~13), and 
$\sim -2.3$ (M~15).} Therefore, we can state
that the any of our M~4/M~13/M~15 targets have essentially 
the same metallicity within the cluster.\\
---  Regarding [K/H], for which values derived from K~{\sc i} 7655 
and 7699 lines are consistent with each other, we notice that 
two stars (M~4/L2406 and M~13/III-73) are evidently anomalous 
because they show considerably discrepant (i.e., larger) [K/H] 
compared to other members. However, when these two stars are 
excluded, we can confirm a reasonable uniformity of 
[K/H] within a cluster; i.e., 
$\langle$[K/H]$_{7665}\rangle$/$\langle$[K/H]$_{7699}\rangle$
($\pm \sigma_{7665}/\pm \sigma_{7699}$) is $-1.01/-1.03$ 
($\pm 0.04/\pm 0.04$) for M~4 (4 stars), $-1.51$/--- ($\pm 0.20$/---)
for M~13 (4 stars), and $-2.26/-2.23$ ($\pm 0.12/\pm 0.09$) 
for M~15 (5 stars). The larger dispersion (0.20) for 
the case of M~13 compared to other two clusters ($\sim$~0.05--0.1) 
may be understood as due to the availability of only one 
K~{\sc i} 7665 line which is severely contaminated by telluric 
lines (cf. figure 2).

\begin{figure}
  \begin{center}
    \FigureFile(75mm,80mm){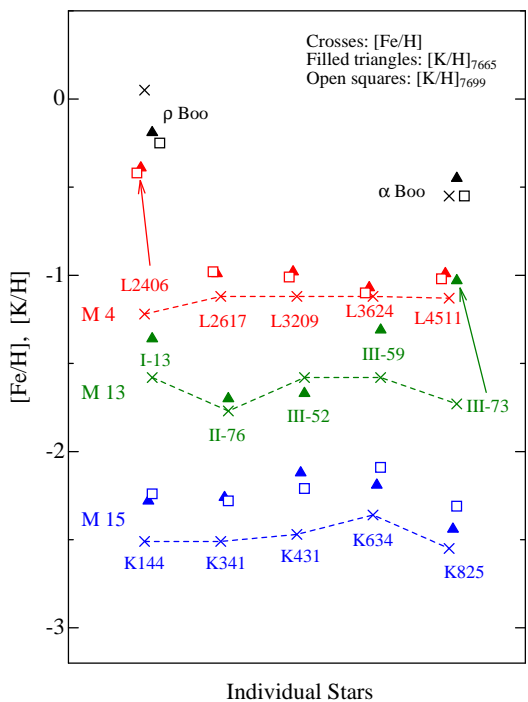}
  \end{center}
\caption{
Graphical plots of the Fe and K abundances (relative to the Sun)
derived for each of the 17 stars (also given in table 1).
The crosses (connected by dashed lines for those 5 stars 
belonging to the same cluster), filled triangles, and
open squares indicate [Fe/H], [K/H]$_{7665}$ (from K~{\sc i} 7665 
line), and [K/H]$_{7699}$ (from K~{\sc i} 7699), respectively.
Plots for M~4, M~13, and M~15 stars are represented in red, 
green, and blue, respectively.
}
\end{figure}

How should the discrepant nature of these two outliers be interpreted?
Actually, we have a good reason to believe that this is nothing but
a superficial effect caused by an inadequate treatment in the
abundance determination (i.e., not the real abundance anomaly).
We point out that the spectra of these M~4/L2406 and M~13/III-73 
have markedly broad line widths (figure 3) and prominently strong
H$\alpha$ emission components (figure 4) than the remaining stars,
which suggests a significant increase in the activity or 
turbulent velocity fields in the upper atmosphere (where the core 
of K~{\sc i} lines form) presumably related to dynamical phenomena 
often seen in AGB stars. We thus suspect the existence of considerably
depth-dependent turbulent velocity dispersion which increases with 
height,\footnote{More generally speaking, such an increasing tendency
of the turbulent velocity field with height is a phenomenon 
occasionally seen in low-gravity stars (though its degree is 
different from case to case); see, e.g., Takeda (1992) for 
$\alpha$~Boo (K giant) or Takeda and Takada-Hidai (1994; appendix B) 
for A--F supergiants.} which may have contributed an additional 
increase to the strength of high-forming saturated K~{\sc i} lines. 
In such a case, an application of the $v_{\rm t}$ value determined 
from Fe lines with $EW < 100$~m$\rm\AA$ would be no more adequate 
for the K~{\sc i} lines with $EW$ of a few $\times$~100~m$\rm\AA$
(especially intensified by the enhanced turbulence in the upper 
atmosphere), for which a somewhat larger $v_{\rm t}$ value should 
have been more relevant. Hence, we believe that the reason for 
the anomalously large [K/H] obtained for these two stars is the 
assignment of inappropriately small $v_{\rm t}$ values, 
which must have resulted in an overestimation of K abundances 
from such strongly saturated K~{\sc i} 7665/7699 lines.

Consequently, regarding that M~4/L2406 and M~13/III-73 correspond 
to exceptional cases of peculiar velocity fields in the upper 
atmospheric layer, we conclude that the abundance uniformity 
essentially holds for [K/H] (as well as for [Fe/H]) within 
M~4, M~13, and M~15.

\subsection{Behavior of [K/Fe] over Wide Metallicities} 

Now that the homogeneity of K (and Fe) abundance within each 
globular cluster has been confirmed, we can regard based on
the argument in section 1 that the observed [K/Fe] ratios (which
naturally turn out also uniform) of the cluster stars retain the 
original composition of the halo gas, from which they formed.

According to table 1 (where the mean of [K/Fe]$_{7665}$ and 
[K/Fe]$_{7699}$ is given), the averaged cluster values of 
$\langle$[K/Fe]$\rangle$ (and the standard deviation $\pm \sigma$) 
are $+0.10 (\pm 0.04)$ (M~4, [Fe/H]~$\simeq -1.1$, 4 stars, 
L2406 excluded), $+0.12 (\pm 0.16)$ (M~13, [Fe/H]~$\simeq -1.7$, 
4 stars, III-73 excluded), and $+0.24 (\pm 0.05)$ (M~15, 
[Fe/H]~$\simeq -2.5$, 5 stars).
We can state from these results that [K/Fe] is marginally 
supersolar at $\sim$~+0.1--0.3 and [K/Fe] tends to
gradually increase with a decrease in the metallicity,
which means a rather tight and monotonic [K/Fe] vs. [Fe/H]
relation (with almost no significant diversity)
in the range of $-2.5 \ltsim$~[Fe/H]~$\ltsim -1$.
These [K/Fe] vs. [Fe/H] relations for 15 stars of 
M~4, M~13, and M~15 (along with $\rho$~Boo and $\alpha$~Boo)
are plotted in figure 6, where the results for nearby F--G disk 
dwarfs ($-1 \ltsim$~[Fe/H]~$\ltsim 0$; taken from Paper I) 
and those for extremely metal-poor stars 
($-4 \ltsim$~[Fe/H]~$\ltsim -2.5$; obtained by reanalyzing 
the $EW$ data of Cayrel et al. 2004; cf. Appendix) are also shown
for comparison. 

\begin{figure}
  \begin{center}
    \FigureFile(75mm,80mm){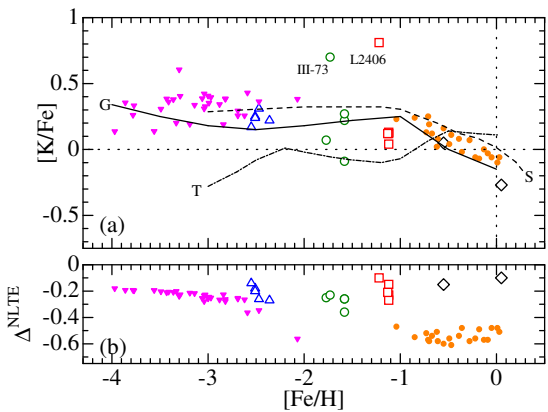}
  \end{center}
\caption{
(a) The [K/Fe] vs. [Fe/H] plots constructed from the 
metallicity and the NLTE potassium abundances given in table 1. 
The results for the 17 program stars are shown in large open symbols
(diamonds for $\rho$~Boo and $\alpha$~Boo, squares for
M~4 stars, circles for M~13 stars, and triangles for M~15 stars). 
The two stars (M~13 III-73 and M~4 L2406) showing exceptionally 
large deviations from the main trend are also indicated.
Meanwhile, the smaller filled symbols represent the results for 
disk stars (circles, taken from Paper I) and extremely metal poor
stars (inverse triangles, reanalysis of Cayrel et al.'s 2004 data,
see the Appendix) presented for comparison.
Lines show the three kinds of theoretical predictions
(see section 1 of Paper I for more details): Dashed line 
(marked ``S'') --- Samland (1998; taken from his figure 14), 
dash-dotted line (marked ``T'') --- Timmes, Woosley, and Weaver (1995; 
the top curve in their figure 24), solid line (marked ``G'') --- 
Goswami and Prantzos (2000; ``Case B'' in their figure 7).
(b) The NLTE corrections for K abundance determinations 
($\Delta^{\rm NLTE}$ is the average of 
$\Delta^{\rm NLTE}_{7665}$ and $\Delta^{\rm NLTE}_{7699}$
in case that both lines are available), plotted against [Fe/H]. 
The same meanings of the symbols as in (a).
}
\end{figure}

As can be clearly seen from this figure, our [K/Fe] results in this 
study for the cluster stars (excluding M~4/L2406 and M~13/III-73) 
and the reference objects ($\rho$~Boo and $\alpha$~Boo) over 
$-2.5 \ltsim$~[Fe/H]~$\ltsim 0$ connect with those of extremely 
metal-deficient stars as well as disk stars quite satisfactorily. 
We may thus conclude that [K/Fe] behaves itself in a rather clear 
manner over a wide [Fe/H] range of $\sim 4$~dex 
from near-solar to ultra-low metallicity. That is, the gradual increase
of [K/Fe] with a decrease in [Fe/H] ($d$[K/Fe]/$d$[Fe/H]~$\sim -0.3$)
seen in disk stars of $-1 \ltsim$~[Fe/H]~$\ltsim 0$ (Paper I) continues 
further down at [Fe/H] $\ltsim -1$ with a slightly decreased slope
until [Fe/H]~$\sim -3$, where [K/Fe] appears to attain a broad maximum    
of $\sim +0.3$ (followed by a sign of weak downturn of [K/Fe] at 
[Fe/H]~$\ltsim -3$). 

Accordingly, we do not regard the considerably large diversity 
of [K/Fe] previously reported for halo stars of [Fe/H]~$\sim -2$ 
(Zhang et al. 2005, Gratton \& Sneden 1987a) to be real, which 
we suspect to be due to some improper treatment in abundance 
determinations (e.g., $EW$ errors caused by unsuccessful removal 
of telluric lines, or use of an inadequate microturbulence such as 
we encountered in the present cases of M~4/L2406 and M~13/III-73).

This observational fact of mildly supersolar [K/Fe] over a wide 
metallicity range (with [K/Fe] $\sim 0$ at near-solar metallicity 
and a broad peak at [Fe/H]~$\sim -3$) has to be explained by
theoretical calculations.
However, theoreticians did not find it easy to reproduce the trend 
of [K/Fe]~$>0$, especially at the very metal-poor regime, because 
the K yield is metallicity-dependent (as a characteristic of 
odd-Z element) and becomes too small there to bring [K/Fe]
above zero. Therefore, comparatively plain calculations predicted
only subsolar (or near-solar at most) [K/Fe] ratios for
metal-poor stars. For example, Timmes, Woosley, and Weaver's (1995) calculations
resulted in a decrease of [K/Fe] ($\ltsim 0$) with a lowering of 
[Fe/H] as shown figure 24 of their paper (also plotted in figure 6
as curve ``T''), which markedly contradicts the observed tendency. 
Therefore, if the supersolar [K/Fe] is to be reenacted even at 
the metal-poor regime, one had to contrive ways to raise it
by some means, such as empirically adjusting the K yield (Samland 1998) 
or using an IMF of larger exponent (cf. Goswami \& Prantzos 2000; 
putting larger weight to comparatively less massive SNeII of lower 
gravitational potential), as described in section 1 of Paper I, 
by which a trend more or less 
similar to the observed one could somehow result
(cf. figure 6 plotted as curves ``S'' and ``G'', respectively). 
At any rate, it is evident that our understanding 
on the synthesis mechanism of K and its evolution in the Galaxy 
is still considerably insufficient (see also the Appendix for
the K yield problem in the extremely metal-poor regime). 
Now that we have become more confident than before about the 
observational [K/Fe] vs. [Fe/H] trend over a wide metallicity 
range, we would encourage theoreticians to revisit this problem,
so that a better match between theory and observation may be 
established in a reasonable manner.

\section{Conclusion}

Given the confusing situation regarding the [K/Fe] ratio
of metal-poor stars in the halo, where some work suggests 
a fairly tight tendency (such like disk stars) while others
report a considerably large diversity amounting to $\sim 1$~dex,
we carried out a spectroscopic study on 15 red giants
of three mildly to very metal-poor globular clusters (M~4, 
M~13, and M~15) along with two reference stars ($\rho$~Boo and 
$\alpha$~Boo) based on the high-dispersion spectra obtained 
with Subaru/HDS, with a purpose of clarifying the behavior 
of [K/Fe] at the relevant metallicity range of 
$-2.5 \ltsim$~[Fe/H]~$\ltsim -1$.

The atmospheric parameters ($T_{\rm eff}$, $\log g$, $v_{\rm t}$,
and [Fe/H]) were spectroscopically determined by using Fe~{\sc i}
and Fe~{\sc ii} lines, and the abundance of K was derived from 
the K~{\sc i} resonance lines at 7664.91 and 7698.97~$\rm\AA$
while taking into the non-LTE effect.

We confirmed that [K/H] (as well as [Fe/H]) is almost homogeneous 
within each of the three clusters to a precision of $\ltsim 0.1$~dex, though superficially large deviations are exceptionally seen for two 
peculiar stars which show signs of considerably increased turbulence 
in the upper atmosphere.

The [K/Fe] ratios of these cluster stars turned out mildly 
supersolar by a few tenths of dex, tending to gradually 
increase from $\sim$~+0.1--0.2 at [Fe/H]~$\sim -1$ to 
$\sim +0.3$ at [Fe/H]~$\sim -2.5$, which is a fairly tight
and clean trend. We thus consider that the previously reported
large diversity of [K/Fe] in halo stars is not real, which 
we suspect to be due to some improper treatment in the analysis.

This result is quite consistent (i.e., smoothly connecting) with 
the [K/Fe] trend of disk stars ($-1 \ltsim$ [Fe/H]) and that of 
extremely metal-poor stars ($-4 \ltsim$~[Fe/H]~$\ltsim -2.5$).
That is, [K/Fe] appears to continue a gradual increase from 
[Fe/H] $\sim 0$ toward a lower metallicity regime down to 
[Fe/H]~$\sim -3$, where a broad maximum of [K/Fe]~$\sim$ +0.3--0.4 
is attained, possibly followed by a slight downturn toward a 
further lower metallicity at [Fe/H]~$\ltsim -3$.

\bigskip

This investigation is based on the data obtained by the 
observation with the Subaru Telescope, which was carried out 
during the practical training of observational astronomy for 
graduate students as a coursework of The Graduate University for 
Advanced Studies (SOKENDAI). 
We express our heartful thanks to S. S. Hayashi, R. Furuya, 
and A. Tajitsu for their continuous support and encouragement, 
as well as to C.-H. Peng and S. Honda for their collaboration 
in the observation.

Helpful comments by N. Tominaga and N. Iwamoto from the theoretical 
side on the first version of the manuscript are also 
acknowledged.

This research made use of the SIMBAD database operated by CDS, 
Strasbourg, France.

\appendix
\section*{Reanalysis of Cayrel et al.'s (2004) Data}

In the extensive spectroscopic study on the sample of 35 extremely 
metal-poor stars toward clarifying the abundance patterns of 17
elements from C to Zn recently carried out by Cayrel et al. (2004), 
the abundances of K were also determined by using K~{\sc i}~7665/7699 
lines. However, they applied a constant non-LTE correction of
$\Delta^{\rm NLTE} = -0.35$~dex to all stars, which they adopted
as a rough estimate based on Ivanova and Shimanski\u{\i}'s (2000)
calculations. Because of this apparently imperfect treatment of 
the non-LTE effect, one can not be sure whether the behavior of [K/Fe]
they obtained (i.e., near-solar or slightly supersolar with 
a decreasing tendency toward a lower metallicity
from [K/Fe]~$\sim +0.2$ at [Fe/H]~$\sim -2.5$ to 
[K/Fe]~$\sim 0$ at [Fe/H]~$\sim -4$; cf. their figure 9) is 
reliable or not.

We decided, therefore, to reanalyze their $EW$ data of 
K~{\sc i}~7665/7699 lines with the atmospheric parameters
they used, while properly taking into account the non-LTE effect. 
Regarding the non-LTE departure coefficients to be included in
line-formation calculations for abundance determination, we used
the already calculated results for a grid of models
($T_{\rm eff}$ from 4500~K to 6500~K, $\log g$ from 1.0 to 5.0,
$v_{\rm t}$ from 1 to 3~km~s$^{-1}$, and [Fe/H] from 0.0 to $-3.0$),
which were originally computed for constructing extensive tables 
of non-LTE corrections\footnote{The anonymous ftp site to access
these electronic tables described in the Appendix of Paper I
is now not available any more. Instead, the same data materials are 
placed at the following web site:
$\langle$http://optik2.mtk.nao.ac.jp/\~{ }takeda/potassium\_nonlte/$\rangle$.}
as explained in the Appendix of Paper I.
Besides, we newly performed calculations corresponding to the case of 
[Fe/H] = $-4$ to be combined with the previous grids, so that we can 
handle any of the relevant 31 stars (for which Cayrel et al. measured 
$EW_{7665}$ and/or $EW_{7665}$) in the parameter range of 
4500~K~$\ltsim T_{\rm eff} \ltsim 5300$~K, 
$0.7 \ltsim \log g \ltsim 2.7$, 
1.2~km~s$^{-1} \ltsim v_{\rm t} \ltsim 2.2$~km~s$^{-1}$,
and $-4.0 \ltsim$~[Fe/H]~$\ltsim -2.1$ by interpolation.
Following the same manner as described in subsection 4.2,
we determined [K/H], $\Delta^{\rm NLTE}$, and [K/Fe] for each star, 
as given in table 4. The resulting values of [K/Fe] 
(as well as $\Delta^{\rm NLTE}$) are plotted against [Fe/H]
in figure 6 (filled inverse triangles). 

As we can see from figure 6b, the extents of the non-LTE corrections 
($|\Delta^{\rm NLTE}|$) systematically decrease with a lowering of
the metallicity; such as from $\Delta^{\rm NLTE} \simeq -0.3$ 
(at [Fe/H] $\simeq -2.5$) to $\Delta^{\rm NLTE} \simeq -0.2$ 
(at [Fe/H] $\simeq -4$), which means that Cayrel et al.'s (2004) 
use of $\Delta^{\rm NLTE} = -0.35$ must have somewhat overcorrected 
(i.e., underestimated) [K/Fe] by $\sim$~0.1--0.2 dex, such that 
the amount of overcorrection progressively increasing with 
a decrease in [Fe/H]. Accordingly, regarding the [K/Fe] vs. [Fe/H] 
relation of extremely metal-poor stars, we consider that 
Cayrel et al.'s (2004) result ([K/Fe] ratio is near-solar or 
slightly above solar ($\sim$~+0.1--0.2) with an appreciable gradient 
of $d$[K/Fe]/$d$[Fe/H]) should be revised as concluded in 
subsection 5.3 (cf. figure 6a): [K/Fe] in this very metal-deficient 
regime still remains well supersolar at $\sim$~+0.2--0.3, with a 
marginal sign of decline (downturn) toward a further lower [Fe/H]. 

The conclusion that [K/Fe] remains supersolar even at [Fe/H]~$\sim -4$
may provide theoreticians with an important constraint on the 
synthesis mechanism of K from the observational side, because
considerable difficulties exist in theoretically reproducing the 
[K/Fe] ratio (cf. subsection 5.3) also in this distinctly low
metallicity regime, where the observed [K/Fe] may be considered
to simply reflect the composition of the first super/hyper-nova ejecta.
For example, Tominaga, Umeda, and Nomoto's (2007) nucleosynthesis calculation 
based on population III supernova models predicted [K/Fe] 
$\sim -1 (\pm 0.5)$ at $-4 \ltsim$~[Fe/H]~$\ltsim -3$ (cf. their 
figure 6), which apparently disagrees with the observational fact
mentioned above. Hence, in order to resolve this situation, a new 
reassessment of assumptions or fundamental physics might as well 
be required.
In this connection, Iwamoto et al.'s (2006) finding, that odd-Z 
elements such as K and Sc can be significantly overproduced
if the proton-rich environment (large initial electron fraction 
$Y_{e}$) is realized, is quite interesting. Further investigation
following this line may be worth a try.

\clearpage
\onecolumn

\setcounter{table}{0}
\setlength{\tabcolsep}{3pt}
\begin{table}[h]
\small
\caption{Stellar parameters and potassium abundance results.}
\begin{center}
\begin{tabular}{c c@{}r@{}c@{}c
r@{}c@{}c r@{}c@{}c c}\hline\hline
Star & $T_{\rm eff}$ & $\log g$ & $v_{\rm t}$ & [Fe/H] & 
$EW_{7665}$ & [K/H]$_{7665}$ & $\Delta_{7665}^{\rm NLTE}$ & 
$EW_{7699}$ & [K/H]$_{7699}$ & $\Delta_{7699}^{\rm NLTE}$ & [K/Fe] \\
  & (K) & (cm~s$^{-1}$) & (km~s$^{-1}$) & (dex) &  
 (m$\rm\AA$) & (dex) & (dex) & (m$\rm\AA$) & (dex) & (dex) & (dex) \\
\hline
$\rho$~Boo & 4363 &  2.09 & 1.29 & +0.05 &  $^{*}$287.7 & $-$0.19 & $-$0.08 &  238.6 & $-$0.25 & $-$0.12 &  $-$0.27 \\
$\alpha$~Boo & 4281 &  1.72 & 1.49 & $-$0.55 &  284.5 & $-$0.45 & $-$0.12 &  235.9 & $-$0.55 & $-$0.19 &  +0.05 \\
\hline
M4~L2406 & 4048 &  0.61 & 2.16 & $-$1.22 &  364.3 & $-$0.39 & $-$0.09 &  319.5 & $-$0.42 & $-$0.12 &  +0.81 \\
M4~L2617 & 4256 &  1.48 & 1.38 & $-$1.12 &  223.1 & $-$0.99 & $-$0.21 &  197.9 & $-$0.98 & $-$0.26 &  +0.13 \\
M4~L3209 & 4025 &  1.13 & 1.49 & $-$1.12 &  255.4 & $-$0.98 & $-$0.13 &  222.5 & $-$1.01 & $-$0.18 &  +0.12 \\
M4~L3624 & 4269 &  1.47 & 1.42 & $-$1.12 &  216.8 & $-$1.07 & $-$0.24 &  189.5 & $-$1.10 & $-$0.30 &  +0.04 \\
M4~L4511 & 4173 &  1.20 & 1.44 & $-$1.13 &  230.8 & $-$0.99 & $-$0.18 &  202.9 & $-$1.02 & $-$0.23 &  +0.12 \\
\hline
M13~I-13 & 4155 &  0.73 & 1.75 & $-$1.58 &  $^{*}$215.4 & $-$1.36 & $-$0.26 & $\cdots$ & $\cdots$ & $\cdots$ &  +0.22 \\
M13~II-76 & 4159 &  0.27 & 1.87 & $-$1.77 &  $^{*}$186.8 & $-$1.70 & $-$0.25 & $\cdots$ & $\cdots$ & $\cdots$ &  +0.07 \\
M13~III-52 & 4271 &  0.80 & 1.55 & $-$1.58 &  $^{*}$168.1 & $-$1.67 & $-$0.36 & $\cdots$ & $\cdots$ & $\cdots$ &  $-$0.09 \\
M13~III-59 & 4206 &  0.54 & 1.42 & $-$1.58 &  $^{*}$190.7 & $-$1.31 & $-$0.26 & $\cdots$ & $\cdots$ & $\cdots$ &  +0.27 \\
M13~III-73 & 4145 &  0.38 & 1.93 & $-$1.73 &  $^{*}$252.8 & $-$1.03 & $-$0.23 & $\cdots$ & $\cdots$ & $\cdots$ &  +0.70 \\
\hline
M15~K144 & 4066 & $-$0.45 & 1.46 & $-$2.51 &  108.2 & $-$2.28 & $-$0.17 &   90.9 & $-$2.24 & $-$0.18 &  +0.25 \\
M15~K341 & 4085 & $-$0.15 & 2.10 & $-$2.51 &  131.2 & $-$2.26 & $-$0.20 &  102.0 & $-$2.28 & $-$0.19 &  +0.24 \\
M15~K431 & 4152 & $-$0.01 & 1.49 & $-$2.47 &  117.8 & $-$2.12 & $-$0.28 &   89.4 & $-$2.21 & $-$0.25 &  +0.31 \\
M15~K634 & 4161 &  0.23 & 1.78 & $-$2.36 &  126.4 & $-$2.19 & $-$0.27 &  111.8 & $-$2.09 & $-$0.27 &  +0.22 \\
M15~K825 & 4012 & $-$0.22 & 1.78 & $-$2.55 &  112.9 & $-$2.44 & $-$0.13 &  101.3 & $-$2.31 & $-$0.15 &  +0.17 \\
\hline
\end{tabular}
\end{center}
Note.\\ 
Columns 2--5 give the atmospheric parameters (the effective 
temperature, the surface gravity, the microturbulence, and 
the Fe abundance relative to the Sun) which were spectroscopically 
determined by using Fe~{\sc i} and Fe~{\sc ii} lines and adopted 
in this study. The line equivalent width, the non-LTE potassium 
abundance relative to the Sun, and the relevant non-LTE correction 
are presented in columns 6--8 (K~{\sc i} 7665 line) and 9--11
(K~{\sc i} 7699 line). In the last column 12 is given the K-to-Fe
logarithmic abundance ratio, [K/Fe] ($\equiv$ [K/H] $-$ [Fe/H]),
where [K/H] is the average of [K/H]$_{7665}$ and [K/H]$_{7699}$
in case both lines are available.\\
$^{*}$ Corrected values for the imperfect removal of telluric lines;
$\Delta EW_{\rm resid}$ corrections of 38.8~m$\rm\AA$ ($\rho$~Boo), 
30.9~m$\rm\AA$ (I-13), 37.9~m$\rm\AA$ (II-76), 47.9~m$\rm\AA$ (III-52), 
72.9~m$\rm\AA$ (III-59), and 22.8~m$\rm\AA$ (III-73) have been 
subtracted from the directly measured $EW$ (cf. footnote 3 in 
subsection 4.1).
\end{table}

\setcounter{table}{1}
\scriptsize
\renewcommand{\arraystretch}{0.8}
\setlength{\tabcolsep}{3pt}
\begin{longtable}{crrcrl}
\caption{Comparison of the atmospheric parameters and metallicity 
with the literature values.}
\hline\hline
Star & $T_{\rm eff}$ & $\log g$ & $v_{\rm t}$ & [Fe/H] & References \\
  & (K) & (cm~s$^{-1}$) & (km~s$^{-1}$) & (dex) &  \\
\hline
\endhead
\hline
\endfoot
\hline
\multicolumn{6}{l}{\hbox to 0pt{\parbox{150mm}{\footnotesize
Note. In case that two kinds of $\log g$ values ($g^{\rm spe}$ and 
$g^{\rm evo}$) are given for the same star (e.g., IVA99 for M~4 stars 
or SNE97 for M~15 stars), we prefereably adopted the spectroscopic 
$\log g^{\rm spe}$.\\
Key to the references: 
BEL85 --- Bell, Edvardsson, and Gustafsson (1985); 
BRO92 --- Brown \& Wallerstein (1992);
CAR00 --- Carr, Sellgren, and Balachandran (2000); 
COH05 --- Cohen \& Melendez (2005);
EDV88 --- Edvardsson (1988); 
FER90 --- Fern\'{a}ndez-Villaca\~{n}as, Rego, and Cornide (1990);
GON98 --- Gonzalez \& Wallerstein (1998); 
GRA82 --- Gratton et al. (1982);
GRA86 --- Gratton \& Ortolani (1986); HIL97 --- Hill (1997);
IVA99 --- Ivans et al. (1999); KRA92 --- Kraft et al. (1992);
KRA93 --- Kraft et al. (1993); KYR86 --- Kyrolainen et al. (1986);
LAM81 --- Lambert \& Ries (1981); 
LEE87 --- Leep, Wallerstein, and Oke (1987);
MCW90 --- McWilliam (1990); MCW94 --- McWilliam \& Rich (1994);
OTS06 --- Otsuki et al. (2006); SHE96 --- Shetrone (1996);
SMI96 --- Smith et al. (1996); SNE91 --- Sneden et al. (1991);
SNE94 --- Sneden et al. (1994); SNE97 --- Sneden et al. (1997);
THE99 --- Th\'{e}venin \& Idiart (1999); 
TOM99 --- Tomkin \& Lambert (1999);
YON08 --- Yong et al. (2008).
}}}
\endlastfoot
\hline
$\rho$~Boo & 4363  & 2.09  & 1.29  & +0.05  & This study (adopted, spectroscopic parameters) \\
 & 4252  & 1.75  & $\cdots$ & 0.00  & This study ($T_{\rm eff}^{\rm pho}$ and $\log g^{\rm evol}$ for reference) \\
 & 4260  & 2.22  & 2.10  & $-$0.17  & MCW90 \\
 \hline
$\alpha$~Boo & 4281  & 1.72  & 1.49  & $-$0.55  & This study (adopted, spectroscopic parameters) \\
 & 4303  & 1.56  & $\cdots$ & $-$0.65  & This study ($T_{\rm eff}^{\rm pho}$ and $\log g^{\rm evol}$ for reference) \\
 & 4490  & 2.01  & 1.80  & $-$0.56  & LAM81 \\
 & 4425  & 1.06  & 2.50  & $-$0.48  & GRA82 \\
 & 4350  & 1.60  & 1.70  & $-$0.58  & BEL85 \\
 & 4330  & 1.50  & 1.50  & $-$0.38  & GRA86 \\
 & 4400  & 1.70  & 2.30  & $-$0.55  & KYR86 \\
 & 4250  & 1.70  & 2.40  & $-$0.60  & LEE87 \\
 & 4375  & 1.97  & 1.80  & $-$0.42  & EDV88 \\
 & 4300  & 2.00  & 1.50  & $-$0.69  & FER90 \\
 & 4280  & 2.19  & 2.30  & $-$0.60  & MCW90 \\
 & 4330  & 2.10  & 1.60  & $-$0.58  & BRO92 \\
 & 4280  & 1.30  & 1.40  & $-$0.54  & MCW94 \\
 & 4300  & 1.50  & 1.70  & $-$0.47  & SNE94 \\
 & 4300  & 1.50  & 1.70  & $-$0.51  & HIL97 \\
 & 4250  & 1.30  & 1.70  & $-$0.68  & GON98 \\
 & 4345  & 2.05  & 1.50  & $-$0.37  & THE99 \\
 & 4300  & 1.50  & 1.70  & $-$0.63  & TOM99 \\
 & 4300  & 1.50  & 1.72  & $-$0.49  & CAR00 \\
 \hline
M4~L2406 & 4048  & 0.61  & 2.16  & $-$1.22  & This study (adopted, spectroscopic parameters) \\
 & 4125  & 0.76  & $\cdots$ & $-$1.23  & This study ($T_{\rm eff}^{\rm pho}$ and $\log g^{\rm evol}$ for reference) \\
 & 4100  & 0.45  & 2.45  & $-$1.20  & IVA99 \\
 & 4150  & 0.15  & 2.20  & $-$1.30  & YON08 \\
 \hline
M4~L2617 & 4256  & 1.48  & 1.38  & $-$1.12  & This study (adopted, spectroscopic parameters) \\
 & 4238  & 1.24  & $\cdots$ & $-$1.21  & This study ($T_{\rm eff}^{\rm pho}$ and $\log g^{\rm evol}$ for reference) \\
 & 4200  & 0.95  & 1.55  & $-$1.17  & IVA99 \\
 & 4275  & 1.25  & 1.65  & $-$1.20  & YON08 \\
 \hline
M4~L3209 & 4025  & 1.13  & 1.49  & $-$1.12  & This study (adopted, spectroscopic parameters) \\
 & 4162  & 0.83  & $\cdots$ & $-$1.33  & This study ($T_{\rm eff}^{\rm pho}$ and $\log g^{\rm evol}$ for reference) \\
 & 3975  & 0.60  & 1.75  & $-$1.20  & IVA99 \\
 & 4075  & 0.75  & 1.95  & $-$1.25  & YON08 \\
 \hline
M4~L3624 & 4269  & 1.47  & 1.42  & $-$1.12  & This study (adopted, spectroscopic parameters) \\
 & 4290  & 1.26  & $\cdots$ & $-$1.22  & This study ($T_{\rm eff}^{\rm pho}$ and $\log g^{\rm evol}$ for reference) \\
 & 4225  & 1.10  & 1.45  & $-$1.16  & IVA99 \\
 & 4225  & 1.05  & 1.60  & $-$1.29  & YON08 \\
 \hline
M4~L4511 & 4173  & 1.20  & 1.44  & $-$1.13  & This study (adopted, spectroscopic parameters) \\
 & 4251  & 1.16  & $\cdots$ & $-$1.21  & This study ($T_{\rm eff}^{\rm pho}$ and $\log g^{\rm evol}$ for reference) \\
 & 4150  & 1.10  & 1.55  & $-$1.19  & IVA99 \\
 & 4150  & 1.05  & 1.70  & $-$1.22  & YON08 \\
 \hline
M13~I-13 & 4155  & 0.73  & 1.75  & $-$1.58  & This study (adopted, spectroscopic parameters) \\
 & 4200  & 0.91  & $\cdots$ & $-$1.57  & This study ($T_{\rm eff}^{\rm pho}$ and $\log g^{\rm evol}$ for reference) \\
 & 4290  & 1.00  & 2.00  & $-$1.46  & KRA93, KRA92, SHE96 \\
 \hline
M13~II-76 & 4159  & 0.27  & 1.87  & $-$1.77  & This study (adopted, spectroscopic parameters) \\
 & 4212  & 0.91  & $\cdots$ & $-$1.65  & This study ($T_{\rm eff}^{\rm pho}$ and $\log g^{\rm evol}$ for reference) \\
 & 4350  & 1.00  & 2.00  & $-$1.49  & KRA93, KRA92, SHE96, SMI96 \\
 & 4300  & 0.85  & 1.95  & $-$1.53  & COH05 \\
 \hline
M13~III-52 & 4271  & 0.80  & 1.55  & $-$1.58  & This study (adopted, spectroscopic parameters) \\
 & 4248  & 1.00  & $\cdots$ & $-$1.56  & This study ($T_{\rm eff}^{\rm pho}$ and $\log g^{\rm evol}$ for reference) \\
 & 4335  & 1.00  & 2.00  & $-$1.50  & KRA93, KRA92, SHE96 \\
 \hline
M13~III-59 & 4206  & 0.54  & 1.42  & $-$1.58  & This study (adopted, spectroscopic parameters) \\
 & 4260  & 0.97  & $\cdots$ & $-$1.50  & This study ($T_{\rm eff}^{\rm pho}$ and $\log g^{\rm evol}$ for reference) \\
 & 4360  & 1.10  & 1.75  & $-$1.45  & KRA93, KRA92, SHE96, SMI96 \\
 \hline
M13~III-73 & 4145  & 0.38  & 1.93  & $-$1.73  & This study (adopted, spectroscopic parameters) \\
 & 4177  & 0.81  & $\cdots$ & $-$1.66  & This study ($T_{\rm eff}^{\rm pho}$ and $\log g^{\rm evol}$ for reference) \\
 & 4300  & 0.85  & 2.25  & $-$1.51  & KRA93, KRA92, SHE96, SMI96 \\
 \hline
M15~K144 & 4066  & $-$0.45  & 1.46  & $-$2.51  & This study (adopted, spectroscopic parameters) \\
 & 4242  & 0.74  & $\cdots$ & $-$2.25  & This study ($T_{\rm eff}^{\rm pho}$ and $\log g^{\rm evol}$ for reference) \\
 & 4390  & 0.90  & 2.00  & $-$2.31  & SNE91 \\
 & 4460  & 0.95  & 2.00  & $-$2.21  & SNE91 \\
 & 4425  & 0.75  & 2.10  & $-$2.34  & SNE97 \\
 \hline
M15~K341 & 4085  & $-$0.15  & 2.10  & $-$2.51  & This study (adopted, spectroscopic parameters) \\
 & 4210  & 0.61  & $\cdots$ & $-$2.35  & This study ($T_{\rm eff}^{\rm pho}$ and $\log g^{\rm evol}$ for reference) \\
 & 4275  & 0.45  & 2.00  & $-$2.34  & SNE97 \\
 \hline
M15~K431 & 4152  & $-$0.01  & 1.49  & $-$2.47  & This study (adopted, spectroscopic parameters) \\
 & 4284  & 0.75  & $\cdots$ & $-$2.30  & This study ($T_{\rm eff}^{\rm pho}$ and $\log g^{\rm evol}$ for reference) \\
 & 4430  & 0.90  & 2.00  & $-$2.28  & SNE91 \\
 & 4500  & 1.00  & 2.00  & $-$2.18  & SNE91 \\
 & 4375  & 0.50  & 2.30  & $-$2.43  & SNE97 \\
 \hline
M15~K634 & 4161  & 0.23  & 1.78  & $-$2.36  & This study (adopted, spectroscopic parameters) \\
 & 4242  & 0.64  & $\cdots$ & $-$2.27  & This study ($T_{\rm eff}^{\rm pho}$ and $\log g^{\rm evol}$ for reference) \\
 & 4225  & 0.30  & 1.85  & $-$2.34  & SNE97 \\
 & 4225  & 0.60  & 2.05  & $-$2.30  & OTS06 \\
 \hline
M15~K825 & 4012  & $-$0.22  & 1.78  & $-$2.55  & This study (adopted, spectroscopic parameters) \\
 & 4242  & 0.63  & $\cdots$ & $-$2.30  & This study ($T_{\rm eff}^{\rm pho}$ and $\log g^{\rm evol}$ for reference) \\
 & 4275  & 0.65  & 1.75  & $-$2.42  & SNE97 \\
\hline
\end{longtable}

\setcounter{table}{2}
\setlength{\tabcolsep}{3pt}
\begin{table}[h]
\small
\caption{Abundance variations in case that the photomeric $T_{\rm eff}$ 
and the evolutionary $\log g$ are used.}
\begin{center}
\begin{tabular}{crcrrrcccc}\hline\hline
Star & $\delta T_{\rm eff}$ & $\delta \log g$  & $\delta$[Fe~{\sc i}/H] & 
$\delta$[Fe~{\sc ii}/H] & $\delta$[Fe/H]$_{\rm av}$ & 
$\delta$[K/H]$_{7665}$ & $\delta$[K/H]$_{7699}$ &
$\delta$[K/H]$_{\rm av}$ & 
$\delta$[K/Fe]$_{\rm av}$ \\
  & (K) & (cm~s$^{-1}$) & (dex) & (dex) & (dex) 
  & (dex) & (dex) & (dex) & (dex) \\
\hline
$\rho$~Boo & $-$111 & $-$0.34 & $-$0.06 & $-$0.04 & $-$0.05 & $-$0.05 & $-$0.06 & $-$0.06 & 0.00 \\
$\alpha$~Boo & +22 & $-$0.16 & $-$0.04 & $-$0.15 & $-$0.10 &+0.07 &+0.05 &+0.06 &+0.16 \\
\hline
M4~L2406 & +77 &+0.15 &+0.03 & $-$0.04 & 0.00 &+0.06 &+0.07 &+0.06 &+0.07 \\
M4~L2617 & $-$18 & $-$0.24 & $-$0.05 & $-$0.13 & $-$0.09 &+0.03 &+0.02 &+0.03 &+0.12 \\
M4~L3209 & +137 & $-$0.30 & $-$0.02 & $-$0.40 & $-$0.21 &+0.25 &+0.22 &+0.24 &+0.44 \\
M4~L3624 & +21 & $-$0.21 & $-$0.02 & $-$0.17 & $-$0.10 &+0.07 &+0.05 &+0.06 &+0.16 \\
M4~L4511 & +78 & $-$0.04 &+0.01 & $-$0.17 & $-$0.08 &+0.11 &+0.11 &+0.11 &+0.19 \\
\hline
M13~I$-$13 & +45 &+0.18 &+0.02 & $-$0.02 & 0.00 &+0.05 & $\cdots$ &+0.05 &+0.05 \\
M13~II$-$76 & +53 &+0.64 &+0.01 &+0.23 &+0.12 & $-$0.02 & $\cdots$ & $-$0.02 & $-$0.14 \\
M13~III$-$52 & $-$23 &+0.20 & $-$0.03 &+0.07 &+0.02 & $-$0.03 & $\cdots$ & $-$0.03 & $-$0.05 \\ 
M13~III$-$59 & +54 &+0.43 &+0.05 &+0.11 &+0.08 &+0.03 & $\cdots$ &+0.03 & $-$0.05 \\
M13~III$-$73 & +32 &+0.43 & 0.00 &+0.15 &+0.07 & $-$0.04 & $\cdots$ & $-$0.04 & $-$0.11 \\
\hline
M15~K144 & +176 & +1.19 &+0.16 &+0.37 &+0.26 &+0.08 &+0.10 &+0.09 & $-$0.17 \\
M15~K341 & +125 &+0.76 &+0.11 &+0.23 &+0.17 &+0.06 &+0.07 &+0.07 & $-$0.10 \\
M15~K431 & +132 &+0.76 &+0.13 &+0.22 &+0.17 &+0.08 &+0.09 &+0.09 & $-$0.09 \\
M15~K634 & +81 &+0.41 &+0.10 &+0.09 &+0.10 &+0.10 &+0.11 &+0.11 &+0.01 \\
M15~K825 & +230 &+0.85 &+0.28 &+0.24 &+0.26 &+0.20 &+0.21 &+0.21 & $-$0.05 \\
\hline
\end{tabular}
\end{center}
Note.\\ 
Given are the abundance variations by using the photometric $T_{\rm eff}$
and evolutionary $\log g$ instead of the spectroscopically
determined ``standard'' parameters adopted in this study (cf. table 1).
(Note that the same $EW$ data for Fe as well as K lines and the 
same microturbulence were used in this test calculation.)
Columns 2 and 3 give the relevant changes in $T_{\rm eff}$ 
($T_{\rm eff}^{\rm pho} - T_{\rm eff}^{\rm spe}$) and $\log g$
($\log g^{\rm evol} - \log g^{\rm spe}$), followed by the variation in 
the mean Fe abundance derived from Fe~{\sc i} lines (column 4), 
that from Fe~{\sc ii} lines (column 5), and the average of both
(column 6). Similarly, the resulting changes in the K abundances
are presented in columns 7 (non-LTE K abundance from the 7665 line),
8 (non-LTE K abundance from the 7699 line), and 9 (the average of both). 
Finally, column 10 gives the variation in [K/Fe], which is simply 
the difference in the results of columns 9 and 6.
\end{table}

\setcounter{table}{3}
\setlength{\tabcolsep}{3pt}
\begin{table}[h]
\small
\caption{Non-LTE reanalysis of Cayrel et al.'s (2004) data.}
\begin{center}
\begin{tabular}{c c r c c
r@{}c@{}c r@{}c@{}c c}\hline\hline
Star & $T_{\rm eff}$ & $\log g$ & $v_{\rm t}$ & [Fe/H] & 
$EW_{7665}$ & [K/H]$_{7665}$ & $\Delta_{7665}^{\rm NLTE}$ & 
$EW_{7699}$ & [K/H]$_{7699}$ & $\Delta_{7699}^{\rm NLTE}$ & [K/Fe] \\
  & (K) & (cm~s$^{-1}$) & (km~s$^{-1}$) & (dex) &  
 (m$\rm\AA$) & (dex) & (dex) & (m$\rm\AA$) & (dex) & (dex) & (dex) \\
\hline
HD~2796      & 4950 &  1.5 &  2.1 & $-$2.47 &  79.8 & $-$2.08 & $-$0.39 &  49.9 & $-$2.13 & $-$0.30 & +0.36 \\
HD~122563    & 4600 &  1.1 &  2.0 & $-$2.82 &  $\cdots$ &$\cdots$ &$\cdots$ &  44.8 & $-$2.44 & $-$0.26 & +0.38 \\
HD~186478    & 4700 &  1.3 &  2.0 & $-$2.59 &  93.0 & $-$2.10 & $-$0.43 &  56.0 & $-$2.22 & $-$0.29 & +0.43 \\
BD+17$^{\circ}$3248  & 5250 &  1.4 &  1.5 & $-$2.07 &  91.0 & $-$1.67 & $-$0.66 &  64.9 & $-$1.71 & $-$0.46 & +0.38 \\
BD$-$18$^{\circ}$5550  & 4750 &  1.4 &  1.8 & $-$3.06 &  41.5 & $-$2.66 & $-$0.27 &  24.0 & $-$2.68 & $-$0.24 & +0.39 \\
CD$-$38$^{\circ}$245   & 4800 &  1.5 &  2.2 & $-$4.19 &  $\cdots$ &$\cdots$ &$\cdots$ &  $\cdots$ &$\cdots$ &$\cdots$ &$\cdots$ \\
BS~16467$-$062 & 5200 &  2.5 &  1.6 & $-$3.77 &   5.6 & $-$3.35 & $-$0.20 &   2.0 & $-$3.52 & $-$0.19 & +0.33 \\
BS~16477$-$003 & 4900 &  1.7 &  1.8 & $-$3.36 &  $\cdots$ &$\cdots$ &$\cdots$ &  10.3 & $-$2.98 & $-$0.21 & +0.38 \\
BS~17569$-$049 & 4700 &  1.2 &  1.9 & $-$2.88 &  50.0 & $-$2.57 & $-$0.28 &  35.0 & $-$2.51 & $-$0.26 & +0.34 \\
CS~22169$-$035 & 4700 &  1.2 &  2.2 & $-$3.04 &  $\cdots$ &$\cdots$ &$\cdots$ &  23.5 & $-$2.73 & $-$0.23 & +0.31 \\
CS~22172$-$002 & 4800 &  1.3 &  2.2 & $-$3.86 &   5.8 & $-$3.61 & $-$0.19 &   4.8 & $-$3.40 & $-$0.20 & +0.35 \\
CS~22186$-$025 & 4900 &  1.5 &  2.0 & $-$3.00 &  39.1 & $-$2.59 & $-$0.26 &  $\cdots$ &$\cdots$ &$\cdots$ & +0.41 \\
CS~22189$-$009 & 4900 &  1.7 &  1.9 & $-$3.49 &  12.5 & $-$3.18 & $-$0.20 &  $\cdots$ &$\cdots$ &$\cdots$ & +0.31 \\
CS~22873$-$055 & 4550 &  0.7 &  2.2 & $-$2.99 &  $\cdots$ &$\cdots$ &$\cdots$ &  33.3 & $-$2.64 & $-$0.24 & +0.35 \\
CS~22873$-$166 & 4550 &  0.9 &  2.1 & $-$2.97 &  $\cdots$ &$\cdots$ &$\cdots$ &  37.5 & $-$2.57 & $-$0.25 & +0.40 \\
CS~22878$-$101 & 4800 &  1.3 &  2.0 & $-$3.25 &  $\cdots$ &$\cdots$ &$\cdots$ &  $\cdots$ &$\cdots$ &$\cdots$ &$\cdots$ \\
CS~22885$-$096 & 5050 &  2.6 &  1.8 & $-$3.78 &   5.0 & $-$3.51 & $-$0.19 &   2.5 & $-$3.53 & $-$0.19 & +0.26 \\
CS~22891$-$209 & 4700 &  1.0 &  2.1 & $-$3.29 &  30.2 & $-$2.88 & $-$0.23 &  16.7 & $-$2.89 & $-$0.22 & +0.41 \\
CS~22892$-$052 & 4850 &  1.6 &  1.9 & $-$3.03 &  $\cdots$ &$\cdots$ &$\cdots$ &  19.8 & $-$2.70 & $-$0.23 & +0.33 \\
CS~22896$-$154 & 5250 &  2.7 &  1.2 & $-$2.69 &  33.1 & $-$2.43 & $-$0.27 &  22.2 & $-$2.37 & $-$0.25 & +0.29 \\
CS~22897$-$008 & 4900 &  1.7 &  2.0 & $-$3.41 &  15.5 & $-$3.08 & $-$0.21 &   9.6 & $-$3.02 & $-$0.21 & +0.36 \\
CS~22948$-$066 & 5100 &  1.8 &  2.0 & $-$3.14 &  $\cdots$ &$\cdots$ &$\cdots$ &  11.2 & $-$2.80 & $-$0.22 & +0.34 \\
CS~22949$-$037 & 4900 &  1.5 &  1.8 & $-$3.97 &   2.8 & $-$3.86 & $-$0.18 &   1.6 & $-$3.81 & $-$0.18 & +0.14 \\
CS~22952$-$015 & 4800 &  1.3 &  2.1 & $-$3.43 &  $\cdots$ &$\cdots$ &$\cdots$ &  10.5 & $-$3.05 & $-$0.21 & +0.38 \\
CS~22953$-$003 & 5100 &  2.3 &  1.7 & $-$2.84 &  $\cdots$ &$\cdots$ &$\cdots$ &  15.6 & $-$2.65 & $-$0.23 & +0.19 \\
CS~22956$-$050 & 4900 &  1.7 &  1.8 & $-$3.33 &  $\cdots$ &$\cdots$ &$\cdots$ &   7.4 & $-$3.13 & $-$0.20 & +0.20 \\
CS~22966$-$057 & 5300 &  2.2 &  1.4 & $-$2.62 &  34.0 & $-$2.38 & $-$0.27 &  21.7 & $-$2.35 & $-$0.26 & +0.26 \\
CS~22968$-$014 & 4850 &  1.7 &  1.9 & $-$3.56 &   6.7 & $-$3.50 & $-$0.19 &   5.1 & $-$3.34 & $-$0.20 & +0.14 \\
CS~29491$-$053 & 4700 &  1.3 &  2.0 & $-$3.04 &  48.1 & $-$2.62 & $-$0.28 &  $\cdots$ &$\cdots$ &$\cdots$ & +0.42 \\
CS~29502$-$041 & 4800 &  1.5 &  1.8 & $-$2.82 &  52.9 & $-$2.46 & $-$0.29 &  36.1 & $-$2.42 & $-$0.26 & +0.38 \\
CS~29502$-$042 & 5100 &  2.5 &  1.5 & $-$3.19 &  13.9 & $-$3.00 & $-$0.21 &  $\cdots$ &$\cdots$ &$\cdots$ & +0.19 \\
CS~29516$-$024 & 4650 &  1.2 &  1.7 & $-$3.06 &  $\cdots$ &$\cdots$ &$\cdots$ &  $\cdots$ &$\cdots$ &$\cdots$ &$\cdots$ \\
CS~29518$-$051 & 5200 &  2.6 &  1.4 & $-$2.69 &  30.0 & $-$2.53 & $-$0.25 &  25.0 & $-$2.34 & $-$0.25 & +0.26 \\
CS~30325$-$094 & 4950 &  2.0 &  1.5 & $-$3.30 &  30.0 & $-$2.69 & $-$0.24 &  $\cdots$ &$\cdots$ &$\cdots$ & +0.61 \\
CS~31082$-$001 & 4825 &  1.5 &  1.8 & $-$2.91 &  $\cdots$ &$\cdots$ &$\cdots$ &  $\cdots$ &$\cdots$ &$\cdots$ &$\cdots$ \\
\hline
\end{tabular}
\end{center}
Note.\\ 
The atmospheric parameters (columns 2--5) as well as the $EW$ values
for the 7665 and 7699 lines (columns 6 and 9) are from Cayrel et al.
(2004). Otherwise, the meanings of the given results are the same
as in table 1.
\end{table}

\end{document}